\documentclass{ws-p8-50x6-00}

\begin{document}

\title{\hfill{\small {\bf MKPH-T-00-10}}\\ \vspace*{.4cm}
The GDH sum rule for the deuteron}

\author{H. Arenh\"ovel}

\address{Institut f\"{u}r Kernphysik, Johannes Gutenberg-Universit\"{a}t,
       D-55099 Mainz, Germany\\E-mail: arenhoevel@kph.uni-mainz.de}

\maketitle

\abstracts{
As first topic, the GDH sume rule is discussed in the context of a more 
general class of sum rules associated with the various contributions to the 
total photoabsorption cross section for target and beam polarization. Then 
I address the question of whether the GDH sum rule for the neutron can be 
determined from the one for the deuteron. It appears that this will not 
be possible in a simple manner. 
The spin response of the deuteron is calculated including 
contributions from the photodisintegration channel and from coherent and 
incoherent single pion production as well, and the GDH integral is 
evaluated up to a photon energy of 550 MeV. The photodisintegration 
channel converges fast enough and gives a large negative contribution, 
essentially from the $^1S_0$ state near threshold and its absolute size is 
about the same than the sum of proton and neutron GDH values. It is only 
partially cancelled by the single pion production contribution. But the 
incoherent channel has not reached convergence at 550 MeV. 
}

\section{Introduction}
The Gerasimov-Drell-Hearn (GDH) sum rule connects the anomalous magnetic 
moment 
of a particle with the energy weighted integral - henceforth denoted by 
$I^{GDH}$ - from threshold up to infinity over the difference of the total 
cross sections for the absorption of circularly polarized photons on a target 
with spin parallel ($\sigma ^P(k)$) and antiparallel ($\sigma ^A(k)$) to 
the spin of the photon. 
In detail it reads for a particle of mass $M_t$, charge $eQ$, anomalous 
magnetic moment $\kappa$ and spin $I$ 
\begin{equation}
I^{GDH}=4\pi^2\kappa^2\frac{e^2}{M_t^2}\,I
=\int_0^\infty \frac{dk}{k}
\left(\sigma ^P(k)-\sigma ^A(k)\right)
\,,\label{GDHSR}
\end{equation}
where the anomalous magnetic moment is 
defined by the total magnetic moment operator of the particle 
$\vec M = (Q+\kappa)\frac{e}{M_t}\vec S$,
where $\vec S$ denotes the spin operator of the target. 

This sum rule gives a very interesting relation between a 
magnetic ground state property of a particle and an integral property of 
its whole excitation spectrum. It shows that the 
existence of a nonvanishing anomalous magnetic moment is directly tied to an 
internal dynamic structure of the particle. Furthermore, it tells us, 
because the lhs of (\ref{GDHSR}) is positive, that the integrated, 
energy-weighted 
total absorption of a circularly polarized photon on a particle with its 
spin parallel to the photon spin is larger than the one on a target with 
its spin antiparallel, if the anomalous magnetic moment does not vanish.

The GDH sum rule has first been derived by Gerasimov\cite{Ger65} and, 
shortly afterwards, independently by Drell and Hearn\cite{DrH66} and also,
less well known, by Hosada and Yamamoto\cite {HoY66}. The last authors 
have used current algebra relations while the others based the derivation 
on the low energy theorem for the Compton scattering amplitude of a 
particle\cite{LET} and the assumption of an 
unsubtracted dispersion 
relation for the difference of the elastic forward scattering
amplitudes for circularly polarized photons and a completely polarized
target with spin parallel and antiparallel to the photon spin. 

First, I will briefly discuss a general class of photoabsorption sum rules of 
which the GDH sume rule is a special case. Then I will address the question 
whether the GDH sume rule of the neutron can be determined from the GDH 
sum rule of the deuteron in the absence of free neutron targets. 
Subsequently, I will present results on an evaluation of the GDH sum 
rule for the deuteron by explicit integration of the GDH integral up to a 
photon energy of 550 MeV including the photodisintegration channel as well 
as coherent and incoherent single pion photoproduction channels. 
I will close with some conclusions and an outlook.

\section{A general class of photoabsorption sum rules}
The GDH sum rule belongs to a larger class of photoabsorption sum rules 
related to the various contributions to the total photoabsorption cross 
section for the general case of beam and target polarization\cite{Are68} 
\begin{eqnarray}
{\sigma_{\mathrm{tot}}}(k,{\rho^\gamma},{\rho^t})
=\frac{1}{2}\,
\sum_{J=0}^{2I}{P_J^t}&[&(1+(-)^J)\,{\sigma_J^{11}(k)}\nonumber\\
&+&(1-(-)^J)\,{P_c^\gamma}\,{\sigma_J^{11}(k)}\,
P_J(\cos{\theta_t})\nonumber\\
&+&(1+(-)^J){P_l^\gamma}\,{\sigma_J^{-11}(k)}
\,d_{20}^J({\theta_t})\,\cos(2{\phi_t})]\,,
\end{eqnarray}
where ${P_l^\gamma}$ and ${P_c^\gamma}$ denote the degree 
of linear and circular photon polarization, respectively.
Furthermore, the target polarization parameters ${P_J^t}$ with respect to an 
orientation direction, characterized by the 
angles ${\theta_t}$ and ${\phi_t}$, are defined by the target 
polarization density matrix
\begin{eqnarray}
\rho^t_{M\, M'}=\frac{(-)^{I-M'}}{\hat I}\sum_{J,m}&\hat J&\,
\left(\matrix{I & I & J\cr M' & -M & -m\cr}\right)\,
{P^t_J}\, e^{im{\phi_t}}\,d^J_{m0}({\theta_t})\,.
\end{eqnarray}
The separate contributions ${\sigma_J^{\lambda'\lambda}}$ are related to 
the forward Compton scattering amplitude via the optical theorem
\begin{eqnarray}
{\sigma_J^{\lambda'\lambda}(k)}=
\frac{4\pi}{k}\,\Im m\,{T^J_{\lambda'\lambda}(k)}\,,
\end{eqnarray}
where $T^J_{\lambda'\lambda}$ is defined by 
\begin{eqnarray}
{T_{\lambda'M',\lambda M}(k)}=
(-)^{I-M}\,\hat I\sum_{J=0}^{2I}\hat J\,
\left(\matrix{I & J & I\cr -M' & \lambda - \lambda' & M\cr}\right)
{T^J_{\lambda'\lambda}(k)}\,.\label{TJ}
\end{eqnarray}
It can be expressed in terms of generalized polarizabilities
\begin{eqnarray}
{T^J_{\lambda'\lambda}(k)}=\frac{\hat J}{\hat I}\,\sum_{L'L}(-)^{L'+L}
\left(\matrix{L & L' & J\cr \lambda & - \lambda' & \lambda'-\lambda\cr}\right)
{P^{L'L\lambda'\lambda}_J(k)}\,
\end{eqnarray}
with
\begin{eqnarray}
{P^{L'L\lambda'\lambda}_J(k)}=\sum_{\nu',\nu=0,1}\lambda^{\prime\nu'}
\,\lambda^{\nu}{P_J(M^{\nu'}L',M^\nu L;k)}\,,
\end{eqnarray}
where $M^0=E$ (electric) and $M^1=M$ (magnetic) multipole. 
The ${T^J_{\lambda'\lambda}}$ are also related to the expansion of the 
scattering amplitude in terms of a complete set of operators 
${\tau^{[J]}}$ with $J=0,1,\dots ,2I$ in the ground state spin space
with reduced matrix elements $\langle I||\tau^{[J]}||I\rangle=\hat I\,\hat J$
\begin{eqnarray}
{T_{\lambda'M',\lambda M}(k)}=\sum_{J=0}^{2I}(-)^{-\lambda'+\lambda}
\langle IM'|{\tau^{[J]}_{\lambda-\lambda'}}|IM\rangle 
{\Omega^{[J]}_{\lambda'-\lambda}(k)}\,.
\end{eqnarray}
Comparison with (\ref{TJ}) leads to the simple relation
\begin{eqnarray}
{T^J_{\lambda'\lambda}(k)}={\Omega^{[J]}_{\lambda'-\lambda}(k)}\,.
\end{eqnarray}
Specifically one has 
\begin{eqnarray}
{\sigma_J^{11}}&=&\frac{\hat J}{\hat I}\,\sum_M (-)^{I-M}\,
\left(\matrix{I & J & I\cr -M & 0 & M\cr}\right)\,
{\sigma_{1M}}\,,
\end{eqnarray}
where ${\sigma_{1M}}$ 
denotes the total cross section for the absorption of a photon with 
helicity $\lambda=1$ by a target with definite spin projection M on the 
photon momentum. Corresponding expressions hold for ${\sigma_J^{-11}}$ 
with respect to the absorption of linearly polarized photons.
In detail one finds for $J=0,1,2$ 
\begin{eqnarray}
\sigma_0^{11}&=&\frac{1}{{\hat I}^2}\,\sum_M \sigma_{1M}\,,\\
\sigma_1^{11}&=&\frac{\sqrt{3}}{{\hat I}^2\,\sqrt{I(I+1)}}\,
\sum_M M\,\sigma_{1M}\,,\\
\sigma_2^{11}&=&\frac{\sqrt{5}}
{{\hat I}^2\,\sqrt{I(I+1)}}\,
\sum_M \frac{(3M^2-I(I+1))}{\sqrt{(2I-1)(2I+3)}}\,\sigma_{1M}\,,
\end{eqnarray}
and for the spin asymmetry ($\sigma^{P/A}=\sigma_{1,\pm I}$)
\begin{eqnarray}
{\sigma^P-\sigma^A}&=&\hat I\,\sum_J \hat J\,(1-(-)^J)\,
\left(\matrix{I & I & J\cr I & -I & 0\cr}\right)\,
{\sigma_J^{11}}\nonumber\\
&=&\frac{2\,\sqrt{3\,I}}{\sqrt{I+1}}\sigma_1^{11}+\cdots\label{spinasy}
\end{eqnarray}
Now, crossing symmetry implies
\begin{eqnarray}
(T^J_{\lambda'\lambda}(-k))^*=(-)^J\,T^J_{\lambda'\lambda}(k)\,.\label{CS}
\end{eqnarray}
For {$J=\mbox{even}$} one takes a once-subtracted dispersion relation for 
$T^J_{\lambda'\lambda}(k)$
\begin{eqnarray}
\Re e\,\Big(T^J_{\lambda'\lambda}(k)-T^J_{\lambda'\lambda}(0)\Big)&=&
\frac{2k^2}{\pi}\,{\cal P}\,
\int_0^\infty \frac{dk'}{k'}\,
\frac{\Im m\,T^J_{\lambda'\lambda}(k')}{k^{\prime 2}-k^2}\nonumber\\
&=&\frac{k^2}{2\pi^2}\,{\cal P}\, \int_0^\infty dk'\,
\frac{\sigma_J^{\lambda'\lambda}(k')}{k^{\prime 2}-k^2}\,,
\end{eqnarray}
while for {$J=\mbox{odd}$} an unsubtracted dispersion relation applies
\begin{eqnarray}
\Re e\,T^J_{\lambda'\lambda}(k)&=&\frac{2k}{\pi}\,{\cal P}\,
\int_0^\infty dk'\,\frac{\Im m\,T^J_{\lambda'\lambda}(k')}{k^{\prime 2}-k^2}
\nonumber\\
&=&\frac{k}{2\pi^2}\,{\cal P}\, \int_0^\infty dk'\,k'
\frac{\sigma_J^{\lambda'\lambda}(k')}{k^{\prime 2}-k^2}\,.
\end{eqnarray}
A power series expansion according to (\ref{CS})
\begin{eqnarray}
\Re e\,T^J_{\lambda'\lambda}(k)=\left\{
\begin{array}{ll}
\sum_{\nu=0}^\infty t_\nu^{\lambda'\lambda,\,J}\,k^\nu & 
\mbox{ for }J\mbox{ even,}\\
\sum_{\nu=0}^\infty t_\nu^{\lambda'\lambda,\,J}\,k^{\nu+1} & 
\mbox{ for }J\mbox{ odd,}\\
\end{array}\right.
\end{eqnarray}
yields a class of sum rules
\begin{eqnarray}
t_\nu^{\lambda'\lambda,\,J}=\left\{
\begin{array}{ll}
\frac{1}{2\pi^2}\,\int_0^\infty dk'\,
\frac{\sigma_J^{\lambda'\lambda}(k')}{k^{\prime 2\nu}} & 
{ \mbox{ for }J\mbox{ even and }\nu=1,2,\dots,}\\
& \\
\frac{1}{2\pi^2}\,\int_0^\infty dk'\,
\frac{\sigma_J^{\lambda'\lambda}(k')}{k^{\prime 2\nu+1}} & 
{ \mbox{ for }J\mbox{ odd and }\nu=0,1,\dots,}\\
\end{array}\right.
\end{eqnarray}
one of which is the GDH, namely for $J=$ odd and $\nu=0$.
Because from the low-energy expansion of the Compton amplitude
\begin{eqnarray}
{T_{\lambda M, \lambda M}(k)}=
{-e^2\,\frac{Q^2}{M_t} +\lambda\,\kappa^2\,
\frac{e^2}{M_t^2}\,\langle S_z \rangle_{IM}\,k} +{\cal O}(k^2)\,,\label{CALE}
\end{eqnarray}
one finds  specifically
\begin{eqnarray}
T^J_{\lambda'\lambda}(k)&=&\left\{
\begin{array}{ll}
{-\delta_{\lambda'\lambda}\,\delta_{J0}\,
e^2\,\frac{Q^2}{M_t}}+{\cal O}(k^2)& 
\mbox{ for }J\mbox{ even,}\\
k\Big({\delta_{\lambda'\lambda}\,\delta_{J1}\,
\lambda\,\kappa^2\,\frac{e^2}{M_t^2}\,\frac{\sqrt{I(I+1)}}{\sqrt{3}}}
+{\cal O}(k^2)\Big)& 
\mbox{ for }J\mbox{ odd.}\\
\end{array}\right.
\end{eqnarray}
The latter yields the GDH sum rule in the form
\begin{eqnarray}
4\,\pi^2\,{\frac{\kappa^2\,e^2}{M_t^2}\,I} &=&
\frac{2\,\sqrt{3I}}{\sqrt{I+1}}\,\int_0^\infty \frac{ dk'}{k'}\,
{\sigma_1^{11}(k')}\,,
\end{eqnarray}
from which (\ref{GDHSR}) follows because of (\ref{spinasy}) and the fact 
that the higher order terms $\sigma_J^{11}$ for $J>1$ do not 
contribute to the left hand side.

\section{Is it possible to get the GDH sum rule for the neutron from the 
one of the deuteron?}

With respect to the GDH sum rule for the neutron, it has been suggested 
to measure in the absence of neutron targets its spin asymmetry using 
a polarized deuteron target. It would rest on two assumptions:

(i) 
A vector polarized deuteron constitutes effectively a polarized 
neutron target.

(ii)
The contribution of the meson production to the spin asymmetry of the 
deuteron is dominated by the quasifree process, and one can neglect binding 
and final state interaction effects arising from the presence of the 
spectator nucleon, so that the deuteron spin asymmetry is an incoherent sum 
of proton and neutron contributions. 

However, I would like to point out a few ``caveats'' which make it very 
unlikely that one can determine the spin asymmetry of the neutron in this way:

(i)
First of all, the neutron is not completely polarized in a completely 
vector polarized deuteron target, i.e., $P(n)=1-1.5\,p_D$,
and its polarization degree is slightly model dependend 
due to the appearance of the deuteron $D$-wave probability $p_D$ which 
is not observable. 

(ii)
A model calculation of the impulse approximation (IA) for the incoherent 
pion photoproduction on the deuteron by R.\ Schmidt et al.\cite{ScA96} 
shows already for the unpolarized cross section that the complete IA is 
not the incoherent sum of proton and neutron contributions. In addition, 
final state interaction and other two-body effects will very likely 
add further complications thus spoiling this simple idea.

(iii)
Since polarization observables show in general a stronger sensitivity to 
small dynamical and coherence effects, the spin asymmetry might be even 
more sensitive to the mentioned disturbing effects.

(iv)
The contribution from coherent $\pi^0$ production on the deuteron is 
non-negligible which certainly is not an incoherent sum of $\pi^0$ production 
on proton and neutron.

Thus all these factors will prohibit a simple determination of the neutron 
spin asymmetry by subtracting from the meson production part of the 
deuteron's spin asymmetry the proton one. That does not mean, that one 
does not learn anything about neutron properties. On the contrary, pion 
photoproduction on the deuteron will provide a very important test for the  
understanding of the production process on the neutron. But this can be 
achieved only in the context of a reliable theoretical model which takes 
into acount all important two-body effects.

\section{The GDH sum rule for the deuteron}
In the case of the deuteron, one finds a very interesting cancellation of 
large contributions. The deuteron has isospin zero, excluding most of the 
contribution of the large nucleon anomalous magnetic moments to its 
magnetic moment, and thus one finds a very small anomalous magnetic moment, 
namely $\kappa_d=-.143$ resulting in a GDH prediction of 
$I^{GDH}_d = 0.65\,\mu$b, which is more than two orders of magnitude smaller 
than the nucleon values. Considering the possible absorption processes, 
one notes first that the incoherent pion production on the deuteron is 
dominated by the quasifree production on the nucleons bound in the deuteron. 
This gives a rough estimate for its contribution to the GDH value, 
namely the sum of the proton and neutron GDH values of 438 $\mu$b. 
Another contribution arises from the coherent $\pi^0$ production channel. 
On the other hand, for the additional photodisintegration channel which 
is the only photoabsorption process below the pion production threshold,
one finds 
at very low energies near threshold a sizeable negative contribution which 
arises from the $M1$-transition to the virtual $^1S_0$ state, because this 
state can only be reached if the spins of photon and deuteron are 
antiparallel, and is forbidden for the parallel situation as has been 
pointed out, for example in Ref.\cite{BaD67}. 

We have evaluated explicitly the finite GDH sum rule for the deuteron by 
integrating up to a photon energy 
of 550 MeV. Three contributions have been included: (i) the 
photodisintegration channel $\gamma d \rightarrow n p$, (ii) the coherent 
pion production $\gamma d \rightarrow \pi^0 d$, and (iii) the 
incoherent pion production $\gamma d \rightarrow \pi N N$. The upper 
integration limit of 550 MeV has been chosen because on the one hand one finds 
sufficient convergence for the photodisintegration channel, while on the other 
hand only single pion photoproduction has been considered, thus limiting the 
applicability of the present theoretical treatment to energies not too far 
above the two pion production threshold as long as significant 
contributions from multipion production cannot be expected. 
I will now discuss the three contributions separately and refer for details 
to Ref.\cite{ArK97}. 

The photodisintegration channel is evaluated within 
the nonrelativistic framework as is described in detail in
Ref.\cite{ArS91} but with inclusion of the most important relativistic 
contributions. Explicitly, all electric and magnetic multipoles up to 
the order $L=4$ are considered which means 
inclusion of the final state interaction in all partial waves up to $j=5$.
For the calculation of the initial deuteron and the final n-p scattering 
wave functions we use the realistic Bonn potential 
(r-space version)\cite{MaH87}. 
In the current operator we distinguish the one-body currents with Siegert 
operators (N), explicit meson exchange contributions (MEC) beyond the
Siegert operators, essentially from $\pi$- and $\rho$-exchange, 
contributions from isobar configurations of the wave functions (IC), 
calculated in the impulse approximation\cite{WeA78}, and leading order 
relativistic contributions (RC) of which the spin-orbit current is 
by far the most dominant part. 

\begin{figure}[t]
\begin{center}
\epsfxsize=24pc 
\epsfbox{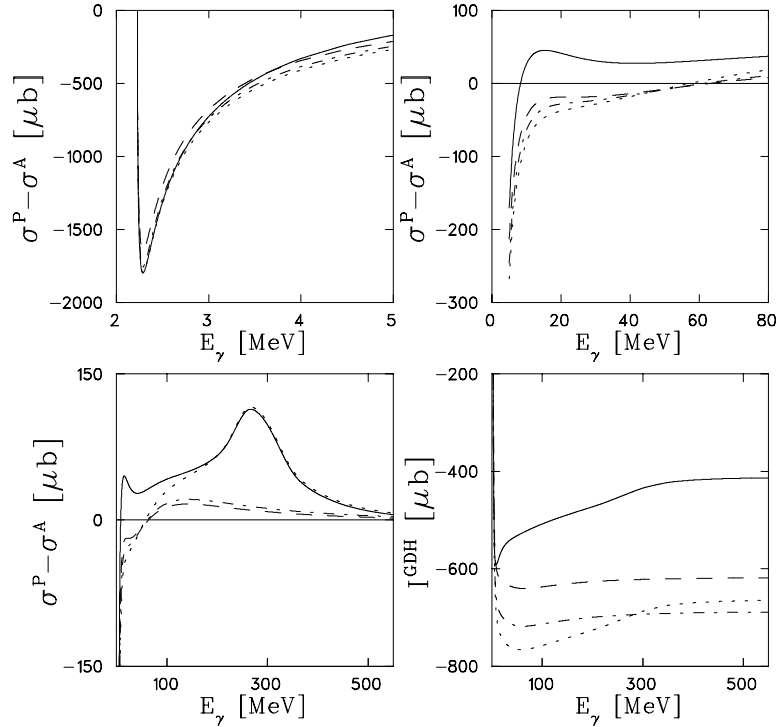} 
\caption{
Contribution of the photodisintegration channel to the GDH sum rule for the 
deuteron. Two upper and lower left panels: difference of the cross sections 
in various energy regions; lower right panel: 
$I^{GDH}_{\gamma d \to np}$ as function of the upper integration energy. 
Dashed curves: N, dash-dot: N+MEC, dotted: N+MEC+IC, and full curves 
N+MEC+IC+RC.}
\label{fig1}
\end{center}
\end{figure}
\begin{table}
\caption{Various contributions of the photodisintegration channel to the 
GDH integral for the deuteron integrated up to 550 MeV in $\mu$b.}
\begin{center}
\footnotesize
 \begin{tabular}{|c|c|c|c|}
\hline
  N & N+MEC & N+MEC+IC & N+MEC+IC+RC\\ 
   \hline
 $-619$ & $-689$ & $-665$ & $-413$ \\
\hline
 \end{tabular}
\label{tabdis}
\end{center}
\end{table}
The results are summarized in Fig.~\ref{fig1}, where the spin asymmetry 
and the GDH integral is shown. The GDH values are listed in 
Tab.~\ref{tabdis}. One readily notes the huge negative contribution from 
the $^1S_0$-state at low energies (see the upper left panel of 
Fig.~\ref{fig1}). Here, the effects from MEC 
are relatively strong, resulting in an enhancement of the negative value by 
about 15 percent. Isobar effects are significant in the region of 
the $\Delta$-resonance, as expected. They give a positive contribution, but 
considerably smaller in absolute size than MEC. The largest positive 
contribution stems from RC in the energy region up to about 100 MeV 
(see the upper right panel of Fig.\ \ref{fig1})
reducing the GDH value in absolute size by more than 30 percent. This 
strong influence from relativistic effects is not surprising in view of the 
fact, that the correct form of the term linear in the photon momentum of 
the low energy expansion of (\ref{CALE}) 
is only obtained if leading order relativistic contributions are included. 
The total sum rule value from the photodisintegration channel then is 
$I^{GDH}_{\gamma d \to np}(550\,\mbox{MeV})=-413\,\mu$b. Its absolute value 
almost equals within less than ten percent the sum of the free proton 
and neutron values. This may not be accidental since the large value is 
directly linked to the nucleon anomalous magnetic moment as is demonstrated 
by the fact that one finds indeed a very small but positive value 
$I^{GDH}_{\gamma d \to np}(550\,\mbox{MeV})=7.3\,\mu$b if the nucleon 
anomalous magnetic moment is switched off in the e.m.\ one-body current 
operator. 

The theoretical model used to calculate the contribution of the
coherent pion production channel is described in detail in 
Ref.\cite{WiA95}. The reaction is clearly dominated by the
magnetic dipole excitation of the $\Delta$ resonance from which one obtains 
a strong positive $I^{GDH}_{\gamma d \to d\pi^0}$ contribution because the 
$\Delta$-excitation is favoured if photon and nucleon spins are parallel. 
The model takes into account pion rescattering by solving a system of 
coupled equations for the N$\Delta$, NN$\pi$ and NN channels. The inclusion 
of the rescattering effects is important and leads in general to a 
significant reduction of the unpolarized cross section in reasonable 
agreement with the differential cross section data available in the 
$\Delta$ region. 
\begin{figure}[t]
\begin{center}
\epsfxsize=20pc 
\epsfbox{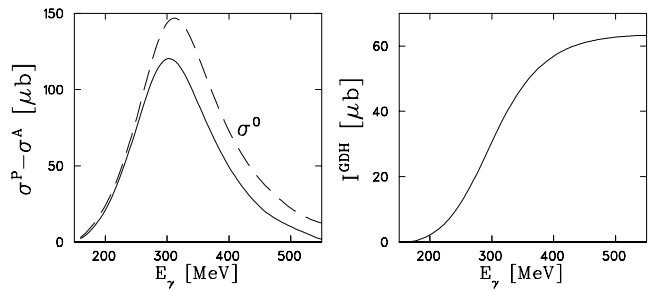} 
\caption{
Contribution of the coherent $\pi^0$ production to the GDH sum rule for the 
deuteron. Left panel: difference of the cross sections (full curve), the 
dashed curve shows the unpolarized cross section; right panel: 
$I^{GDH}_{\gamma d \to d\pi^0}$ as function of the upper integration energy.}
\label{fig2}
\end{center}
\end{figure}
Fig.~\ref{fig2} shows the result of our calculation. One sees the strong 
positive contribution from the $\Delta$-excitation giving a value 
$I^{GDH}_{\gamma d \to d\pi^0}(550\,\mbox{MeV})=63\,\mu$b. The comparison 
with the unpolarized cross section, also plotted in Fig.~\ref{fig2}, 
demonstrates the dominance of $\sigma^P$ over $\sigma^A$. 
Furthermore, one notes quite satisfactory convergence. 

The calculation of the $\gamma d \rightarrow \pi NN$ contributions to
the GDH integral is based on the spectator nucleon approach discussed 
in Ref.\cite{ScA96}. In this framework, the reaction proceeds 
via pion production on one nucleon while the other nucleon acts 
merely as a spectator. Thus, the $\gamma d \rightarrow \pi NN$
operator is given as the sum of the elementary $\gamma N \rightarrow \pi N$ 
operators of the two nucleons. For this elementary operator,  
we have taken the standard pseudovector 
Born terms and the contribution of the $\Delta$ resonance, and a satisfactory 
description of pion photoproduction on the nucleon is achieved in the 
$\Delta$-resonance region\cite{ScA96}. Although the spectator model 
does not include any final state interaction except for the resonant 
$M_{1+}^{3/2}$ multipole, it gives quite a good description of available 
data on the total cross section demonstrating the dominance of the 
quasifree production process, for which the spectator model should work 
quite well. 
\begin{figure}[t]
\begin{center}
\epsfxsize=26pc 
\epsfbox{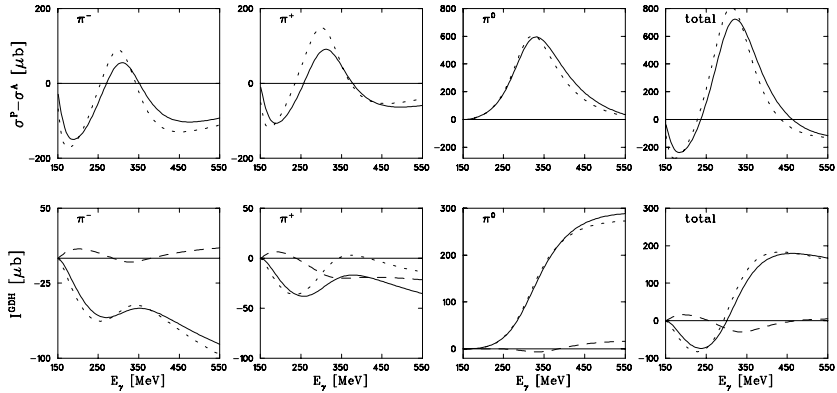} 
\caption{
Contribution of the incoherent $\pi$ production to the GDH sum rule for the 
deuteron and the nucleon. Upper part: difference of the cross sections; 
lower part: $I^{GDH}_{\gamma d \to NN\pi}$ as function of the upper integration energy. Full 
curves for the deuteron, dotted curves for the nucleon. In the case of $\pi^0$ 
production, the dotted curve shows the summed proton and neutron 
contributions.}
\label{fig3}
\end{center}
\end{figure}
The results are collected in Fig.~\ref{fig3}. The upper part shows the 
individual contributions from the different charge states of the pion 
and their total sum to 
the cross section difference for pion photoproduction on both the deuteron 
and for comparison on the nucleon. One notes qualitatively a similar 
behaviour although the maxima and minima are smaller and also slightly 
shifted towards higher energies for the deuteron. In the lower part of 
Fig.~\ref{fig3} the corresponding GDH integrals are shown. A large positive 
contribution comes from $\pi^0$-production whereas the charged pions give a 
negative but - in absolute size - smaller contribution to the GDH value. Up to 
an energy of 550 MeV one finds for the total contribution of the incoherent 
pion production channels a value 
$I^{GDH}_{\gamma d \to NN\pi}(550\,\mbox{MeV})=167\,\mu$b which is remarkably 
close to the sum of the neutron and proton values for the given elementary 
model $I^{GDH}_n(550\,\mbox{MeV})+I^{GDH}_p(550\,\mbox{MeV})=163\,\mu$b. 
However, as is evident from Fig.~\ref{fig3}, convergence is certainly not 
reached at this energy. Thus it is not clear whether this result is accidental,
and in addition, one has to see what the effect of neglected rescattering and 
other two-body contributions will be. Furthermore, the elementary pion 
production operator used in this work had been 
constructed primarily to give a realistic description of the $\Delta$ 
resonance region. In fact, it underestimates the GDH inegral up to 550 MeV 
by about a factor two compared to a corresponding evaluation based on a 
multipole analysis of experimental pion photoproduction data. 

The contributions from all three channels and their sum are listed in 
Tab.~\ref{tab1}. A very interesting and important result is the large negative 
contribution from the photodisintegration channel near and not too far 
above the break-up threshold with surprisingly large relativistic effects 
below 100 MeV. Hopefully, this low energy feature of 
the spin asymmetry can be checked experimentally in the near future. 
\begin{table}
\caption{Contributions of the different absorption channels to the 
GDH integral for the deuteron integrated up to 550 MeV in $\mu$b.}
\begin{center}
\footnotesize
 \begin{tabular}{|c|c|c|c|c|c|}
\hline
   $\gamma d \to np$ & $\gamma  d \to d \pi^0$ & $\gamma d \to np\pi^0$ 
  &$\gamma d \to nn\pi^+$   & $\gamma d \to pp\pi^-$   &total\\
\hline
  $-413$ &   63 &  288 &  $-35$ &  $-86$ & $-183$ \\
\hline
 \end{tabular}
\label{tab1}
 \end{center}
\end{table}
For the total GDH value from explicit integration up to 550 MeV, we find a 
negative value $I^{GDH}_d(550\,\mbox{MeV})=-183\,\mu$b. However, as I have 
mentioned above, some uncertainty lies in the contribution of the incoherent
pion production channel because of shortcomings of the model of the
elementary production amplitude above the $\Delta$ resonance. If one uses 
instead of the model value $I^{GDH}_{\gamma d \to NN\pi}(550\,\mbox{MeV})=
167\,\mu$b the sum of the GDH values of neutron and proton by integrating the 
cross section difference obtained from a multipole analysis of experimental 
data (fit SM95 from the VPI-analysis), giving $I^{GDH}_n(550\,\mbox{MeV})
+I^{GDH}_p(550\,\mbox{MeV})=331\,\mu$b, one finds for the deuteron 
$I^{GDH}_d(550\,\mbox{MeV})=-19\,\mu$b, which I consider a more realistic 
estimate. Since this value is still negative, a positive contribution of 
about the same size should come from contributions at higher energies in 
order to fulfil the small GDH sum rule for the deuteron, provided that the 
sum rule is valid. These contributions should come from the incoherent 
single pion production above 550 MeV, because for this channel convergence 
had not been reached in contrast to the other two channels, and in addition, 
from multipion production. 

\section{Conclusions and Outlook}
In order to summarize let me draw a few important conclusions:
The spin asymmetry of the deuteron is a very interesting observable of its 
own value because of a strong anticorrelation of photodisintegration and 
pion production. It is also very sensitive to relativistic 
effects at quite low energies which have never been tested in detail in this 
observable. 

It is very doubtful, if not impossible, that one can extract in a simple 
manner the neutron spin asymmetry from the spin asymmetry of the deuteron. 
However, the spin asymmetry of the deuteron will provide more detailed tests 
of our understanding of pion photoproduction in a nucleus, and in particular 
on the neutron. 

Future theoretical and experimental studies should be devoted to the 
following topics:

(a) Theory: Improvement of the elementary pion photoproduction amplitude 
above the two pion threshold and inclusion of multiple pion production.
For photodisintegration the inclusion of a retarded $\Delta N$ interaction, 
of higher nucleon resonances, and of relativistic contributions is needed.
Furthermore, other sum rules, alluded to in Sect.~2, should be studied.

(b) Experiment:  A careful measurement of the spin asymmetry for the proton 
over a larger energy range. With respect to the deuteron, a measurement of 
the spin asymmetry is needed, separately for the photodisintegration channel, 
in particular close to the break-up threshold, as well as for the meson 
production channel.

\end{document}